\documentclass[10pt]{article}

\usepackage{amsmath}
\usepackage{amssymb}

\usepackage{graphicx}

\usepackage{cite}

\usepackage{color} 

\usepackage[labelfont=bf,labelsep=period,justification=raggedright]{caption}

\makeatletter
\renewcommand{\@biblabel}[1]{\quad#1.}
\makeatother

\date{}

\newcommand{\be}{\begin{equation}}
\newcommand{\ee}{\end{equation}}
\newcommand{\bea}{\begin{eqnarray}}
\newcommand{\eea}{\end{eqnarray}}

\newcommand{\rme}{{\rm{e}}}
\newcommand{\rmi}{{\rm{i}}}

\begin{document}

% Title must be 150 characters or less
\begin{flushleft}
{\Large
\textbf{A Precise Error Bound for Quantum Phase Estimation}
}
% Insert Author names, affiliations and corresponding author email.
\\
James M.~Chappell$^{1,\ast}$, 
Max A.~Lohe$^{2}$, 
Lorenz von Smekal$^{3}$, 
Azhar Iqbal$^{4}$, 
Derek Abbott$^{5}$
\\
\bf{1,2} School of Chemistry and Physics, University of Adelaide, SA 5005, Australia
\\
\bf{3}  Institut f\"ur Kernphysik, Technische Universit\"at Darmstadt, Schlossgartenstra{\ss}e 9, 64289 Darmstadt, Germany
\\
\bf{4,5} School of Electrical and Electronic Engineering, University of Adelaide, SA 5005, Australia
\\
$\ast$ E-mail: james.m.chappell@adelaide.edu.au
\end{flushleft}

% Please keep the abstract between 250 and 300 words
\section*{Abstract}
Quantum phase estimation is one of the key algorithms in the field of quantum computing, but up until now, only approximate expressions have been derived for the probability of error.  We revisit these derivations, and find that by ensuring symmetry in the error definitions, an exact formula can be found. This new approach may also have value in solving other related problems in quantum computing, where an expected error is calculated.  Expressions for two special cases of the formula are also developed, in the limit as the number of qubits in the quantum computer approaches infinity and in the limit as the extra added qubits to improve reliability goes to infinity. It is found that this formula is useful in validating computer simulations of the phase estimation procedure and in avoiding the  overestimation of the number of qubits required in order to achieve a given reliability. This formula thus brings improved precision in the design of quantum computers.

% Please keep the Author Summary between 150 and 200 words
% Use first person. PLoS ONE authors please skip this step. 
% Author Summary not valid for PLoS ONE submissions.   
%\section*{Author Summary}

\section*{Introduction}

Phase estimation is an integral part of Shor's algorithm \cite{ShorFinal} as well as many other quantum algorithms \cite{NielsenChuang:2002}, designed to run on a quantum computer, and so an exact expression for the maximum probability of error is valuable, in order to efficiently achieve a predetermined accuracy.  
Suppose we wish to determine a phase angle $ \phi $ to an accuracy of $ s $ bits, which hence could be in error, with regard to the true value of $ \phi $, by up to ${ 2^{-s} }$, then due to the probabilistic nature of quantum computers, to achieve this we will need to add $ p $ extra qubits to the quantum register in order to succeed with a probability of $ 1 - \epsilon $.  Quantum registers behave like classical registers upon measurement, returning a one or a zero from each qubit.
Previously, Cleve et al. \cite{CEMM:98} determined the
following upper bound: 
\be \label{eq:transformState97}
p_{\rm{C}} = \left \lceil \log_2 \left (\frac{1}{2 \epsilon} +\frac{1}{2} \right ) \right \rceil . 
\ee
Thus the more confident we wish to be (a small $ \epsilon $), for the output to achieve a given precision $s$, the more qubits, $p$, will need to be added to the quantum register.
Formulas of essentially the same functional form as Eq.~(\ref{eq:transformState97}), are produced by two other authors, in \cite{NielsenChuang:2002} and \cite{IB:2002}, due to the use of similar approximations in their derivation.  For example, we have $ p = \left \lceil \log_2 \left (\frac{1}{2 \epsilon} +2 \right ) + \log_2 \pi \right \rceil $, given in \cite{IB:2002}.
As we show in the following, these approximate error formulas are unsatisfactory in that they overestimate the number of qubits required in order to achieve a given reliability. 

The phase angle is defined as follows, given a unitary operator $ U $, we produce the eigenvalue equation $ U | u \rangle = \rme^{2 \pi \rmi
  \phi}  | u \rangle $, for some eigenvector $ | u \rangle $, and we seek to determine the phase $ \phi \in [0,1)$ 
  using the quantum phase estimation procedure \cite{Mosca:1999}.
  The first stage in phase estimation produces, in the
  measurement register with a $t$ qubit basis $\{ |k\rangle\}$, 
  the state \cite{NielsenChuang:2002}  
  \be  \label{eq:phasesum}
| \tilde\phi \rangle_{\rm{Stage 1}} = \frac{1}{2^{t/2}} \sum_{k=0}^{2^t-1}
  \rme^{2 \pi \rmi \phi k} \, | k \rangle .
\ee
If $\phi = b/2^t$ for some integer $b = 0, \, 1\, ,\dots 2^t-1$,  
then 
\be |\tilde\phi\rangle_\mathrm{Stage 1} =   \sum_{k=0}^{2^t-1} y_k
  |k\rangle 
  \; , \;\; \mbox{with} \;\;\; y_k = \frac{ \rme^{2 \pi \rmi b
  k/2^t}}{2^{t/2}}   ,
\ee 
is the discrete Fourier transform of the basis state $|b\rangle $, that is, the state with amplitudes $x_k = \delta_{kb}$. We then
  read off the exact phase $\phi = b/2^t$ from the inverse Fourier
  transform as $|b\rangle  = \mathcal F^\dagger |\tilde\phi\rangle $.  

In general however, when  $ \phi $ cannot be written in an exact $ t $ bit
binary expansion, the inverse Fourier transform in the final stage of
the phase estimation procedure yields a state 
\be
| \phi \rangle \equiv \cal{F}^{\dagger} | \tilde\phi \rangle_{\rm{Stage
    1}}\; ,
\ee
from which we only obtain an estimate for $ \phi $.
That is, the coefficients $x_k$ of the state $|\phi\rangle $
in the $t$ qubit basis $\{ |k\rangle\}$ will yield probabilities which 
peak at the values of $k$ closest to $\phi $.

Our goal now is to derive an upper bound which avoids the
approximations used in the above formulas and hence obtain a precise
result. 

% Results and Discussion can be combined.
\section*{Analysis}

In order to derive an improved accuracy formula for phase estimation, we initially follow the procedure given in \cite{CEMM:98}, where it is noted, that because of the limited resolution provided by the quantum register of $ t $ qubits, the phase $ \phi $ must be approximated by the fraction $ \frac{b}{2^t} $, where $ b $
is an integer in the range $ 0 $ to $ 2^t -1 $ such that $ b/2^t =
0.b_1 \dots b_t $ is the best $ t $ bit approximation to ${ \phi }$,
which is less than $ \phi $. We then define 
\[
\delta = \phi - b/2^t, 
\]
which is the difference between $ \phi $ and $ b/2^t $ and where
clearly $ 0 \le \delta < 2^{-t} $. 
The first stage of the phase estimation procedure produces the state given by Eq.~(\ref{eq:phasesum}).
Applying the inverse quantum Fourier transform to this state produces
\begin{equation} \label{eq:transformState}
 |\phi \rangle = \sum_{k=0}^{2^t - 1} x_k \, |k\rangle \; ,
\end{equation}
where
\begin{equation} 
x_k =  \frac{1}{2^{t}} \, \sum_{\ell=0}^{2^t - 1}
\rme^{2 \pi \rmi (\phi -   k /2^t ) \ell  } =  \frac{1}{2^{t}} \, 
\frac{1-\rme^{2\pi \rmi \, 2^t\delta} }{1- \rme^{2\pi\rmi (\delta -
    \frac{k-b}{2^t})}} \; . \label{eq:transformCoeff}
\end{equation}

Assuming the outcome of the final measurement is $ m $, we can bound
the probability of obtaining a value of $ m $ such that $ |m-b| \le e
$, where $ e $ is a positive integer characterizing our desired
tolerance to error, where $ m $ and $ b $ are integers such that $ 0
\le m <2^t $ and $ 0 \le b <2^t $.  
The probability of observing such an $ m $ is given by
\begin{equation} \label{eq:transformState6}
pr(|m-b| \le e) = \sum_{\ell=-e }^{e} |x_{b+\ell} |^2 .
\end{equation}
This is simply the sum of the probabilities of the states within $ e $ of $ b $, where
\be \label{eq:alphaEll}
x_{b+\ell} = \frac{1}{2^{t}} \, \frac{1-\rme^{ 2 \pi \rmi \, 2^t \delta }}{1-\rme^{ 2 \pi \rmi ( \delta - \ell/2^t) }}, 
\ee
which is the standard result obtained from Eq.~(\ref{eq:transformCoeff}), in particular see equation 5.26 in \cite{NielsenChuang:2002}.
Typically at this point approximations are now made to simplify $ x_{\ell} $, however we proceed without approximations. We have  
\be
|x_{b+\ell}|^2 =  \frac{1}{2^{2 t}}\,  \frac{1- \cos ( 2 \pi 2^t \delta )
}{1- \cos(2 \pi ( \delta - \ell/2^t)) }\; . 
\ee
Suppose we wish to approximate $ \phi $ to an accuracy of $ 2^{-s} $, that is, we choose $ e = 2^{t-s-1} = 2^{p-1} $, using $ t = s + p $, and if we denote the probability of failure \footnote{Nielsen and Chuang \cite{NielsenChuang:2002} in the preliminary to Eq. 5.35, appear to have written incorrectly $2^{p}-1$ instead of $2^{p-1}$.}
\be \label{eq:defineEps}
\epsilon = p(|m-b|>e) ,
\ee
then we have
\be \label{eq:transformState16}
\epsilon = 1 - \frac{1-\cos 2 \pi 2^t \delta }{2^{2 t}} \sum_{\ell=-2^{p-1} }^{2^{p-1}}  \frac{1}{1- \cos 2 \pi ( \delta - \ell/2^t ) } .
\ee
This formula assumes that for a measurement $ m $, we have a successful result if we measure a state either side of $ b $ within a distance of $ e $, which is the conventional assumption.  

This definition of error however is asymmetric because there will be unequal numbers of states summed about the phase angle $ \phi $ to give the probability of a successful result, because an odd number of states is being summed. We now present a definition of the error which is symmetric about $ \phi $. 

\subsection*{Modified definition of error}

Given an actual angle $ \phi $ that we are seeking to approximate in the phase estimation procedure, a measurement is called successful if it lies within a certain tolerance $ e $ of the true value $ \phi $.  That is, for a measurement of state $ m $ out of a possible $ 2^t $ states, the probability of failure will be
\be \label{eq:errDef}
\epsilon = p \left ( \left | 2 \pi \frac{m}{2^t}-\phi \right | >\frac{1}{2} \frac{2 \pi}{2^s} \right ).
\ee
Thus we consider the angle to be successfully measured accurate to $ s $ bits, if the estimated $ \phi $ lies in the range $ \phi \pm \frac{1}{2} \frac{2 \pi}{2^s} $.
Considering our previous definition Eq.~(\ref{eq:defineEps}), due to the fact that $ b $ is defined to be always less than $ \phi $, then compared to the previous definition of $ \epsilon $, we lose the outermost state at the lower end of the summation in Eq.~(\ref{eq:transformState16}) as shown in Fig.~(\ref{solutionAcc}). For example for $ p = 1 $, the upper bracket in Fig.~(\ref{solutionAcc}) (representing the error bound) can only cover two states instead of three, and so the sum in Eq.~(\ref{eq:transformState16}) will now sum from 0 to 1, instead of $-$1 to 1, for this case.

%\begin{figure}[!ht]
%\begin{center}
%%\includegraphics[width=4in]{figure_name.2.eps}
%\end{center}

\setlength{\unitlength}{1mm}

\begin{figure}[!ht]
\begin{center}
\begin{picture}(85,54)
\put(43.5,1){\vector(0,1){48}} 

\put(10,35) {\line(1,0){65}} 
\put(15,35) {\line(0,1){5}} 
\put(23,35) {\line(0,1){5}} 
\put(31,35) {\line(0,1){5}} 
\put(39,35) {\line(0,1){5}} 
\put(47,35) {\line(0,1){5}} 
\put(55,35) {\line(0,1){5}} 
\put(63,35) {\line(0,1){5}} 
\put(19,35) {\line(0,1){4}} 
\put(27,35) {\line(0,1){4}} 
\put(35,35) {\line(0,1){4}} 
\put(43,35) {\line(0,1){4}} 
\put(51,35) {\line(0,1){4}} 
\put(59,35) {\line(0,1){4}} 
\put(68,37){$ p = 1 $ , $ \sum_0^1 $}  
\put(39.45,42){ $ \overbrace{       \quad          } $ }
\put(42.4,41){$ _0 $}  
\put(46.4,41){$ _1 $}

\put(10,20) {\line(1,0){65}} 
\put(15,20) {\line(0,1){5}} 
\put(23,20) {\line(0,1){5}} 
\put(31,20) {\line(0,1){5}} 
\put(39,20) {\line(0,1){5}} 
\put(47,20) {\line(0,1){5}} 
\put(55,20) {\line(0,1){5}} 
\put(63,20) {\line(0,1){5}} 
\put(19,20) {\line(0,1){4}} 
\put(27,20) {\line(0,1){4}} 
\put(35,20) {\line(0,1){4}} 
\put(43,20) {\line(0,1){4}} 
\put(51,20) {\line(0,1){4}} 
\put(59,20) {\line(0,1){4}} 
\put(17,20) {\line(0,1){3}} 
\put(25,20) {\line(0,1){3}} 
\put(33,20) {\line(0,1){3}} 
\put(41,20) {\line(0,1){3}} 
\put(49,20) {\line(0,1){3}} 
\put(57,20) {\line(0,1){3}} 
\put(21,20) {\line(0,1){3}} 
\put(29,20) {\line(0,1){3}} 
\put(37,20) {\line(0,1){3}} 
\put(45,20) {\line(0,1){3}} 
\put(53,20) {\line(0,1){3}} 
\put(61,20) {\line(0,1){3}} 
\put(39.45,27){ $ \overbrace{       \quad          } $ }
\put(68,22){$ p = 2 $ , $ \sum_{-1}^2 $}  
\put(38.3,25.6){$ _{-1} $}  
\put(42.4,25.6){$ _0 $}  
\put(44.3,25.6){$ _1 $}  
\put(46.5,25.6){$ _2 $}  

\put(10,5) {\line(1,0){65}} 
\put(15,5) {\line(0,1){5}} 
\put(23,5) {\line(0,1){5}} 
\put(31,5) {\line(0,1){5}} 
\put(39,5) {\line(0,1){5}} 
\put(47,5) {\line(0,1){5}} 
\put(55,5) {\line(0,1){5}} 
\put(63,5) {\line(0,1){5}} 
\put(19,5) {\line(0,1){4}} 
\put(27,5) {\line(0,1){4}} 
\put(35,5) {\line(0,1){4}} 
\put(43,5) {\line(0,1){4}} 
\put(51,5) {\line(0,1){4}} 
\put(59,5) {\line(0,1){4}} 
\put(17,5) {\line(0,1){3}} 
\put(25,5) {\line(0,1){3}} 
\put(33,5) {\line(0,1){3}} 
\put(41,5) {\line(0,1){3}} 
\put(49,5) {\line(0,1){3}} 
\put(57,5) {\line(0,1){3}} 
\put(21,5) {\line(0,1){3}} 
\put(29,5) {\line(0,1){3}} 
\put(37,5) {\line(0,1){3}} 
\put(45,5) {\line(0,1){3}} 
\put(53,5) {\line(0,1){3}} 
\put(61,5) {\line(0,1){3}} 
\put(16,5) {\line(0,1){2}} 
\put(24,5) {\line(0,1){2}} 
\put(32,5) {\line(0,1){2}} 
\put(40,5) {\line(0,1){2}} 
\put(48,5) {\line(0,1){2}} 
\put(56,5) {\line(0,1){2}} 
\put(20,5) {\line(0,1){2}} 
\put(28,5) {\line(0,1){2}} 
\put(36,5) {\line(0,1){2}} 
\put(44,5) {\line(0,1){2}} 
\put(52,5) {\line(0,1){2}} 
\put(60,5) {\line(0,1){2}} 
\put(18,5) {\line(0,1){2}} 
\put(26,5) {\line(0,1){2}} 
\put(34,5) {\line(0,1){2}} 
\put(42,5) {\line(0,1){2}} 
\put(50,5) {\line(0,1){2}} 
\put(58,5) {\line(0,1){2}} 
\put(22,5) {\line(0,1){2}} 
\put(30,5) {\line(0,1){2}} 
\put(38,5) {\line(0,1){2}} 
\put(46,5) {\line(0,1){2}} 
\put(54,5) {\line(0,1){2}} 
\put(62,5) {\line(0,1){2}} 
\put(39.45,12){ $ \overbrace{       \quad          } $ }
\put(68,7){$ p = 3 $ , $ \sum_{-3}^4 $}  
\put(38.4,10.5){$ _{-2} $}  
\put(42.4,10.5){$ _0 $}  
\put(44.4,10.5){$ _2 $}  
\put(46.5,10.5){$ _4 $}

\thicklines 
\put(42.5,51){$ \phi $}

\end{picture}
\end{center}

\caption{
{\bf Defining the limits of summation for the phase estimation error. }
For the cases $ p =1,2,3 $, we show the measurements which are accepted as lying within the required distance of $ \phi$, shown by the vertical arrow, which define the limits of summation used in Eq.~(\ref{eq:transformState16Corr}).
}

\label{solutionAcc}
\end{figure}
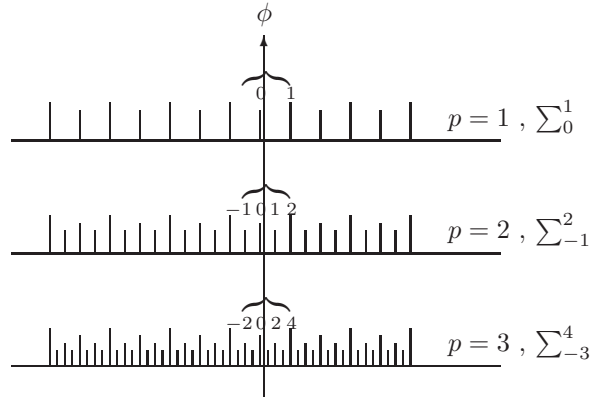

\subsection*{An optimal bound}

Based on this new definition then for all cases we need to add 1 to the lower end of the summation giving
\be \label{eq:transformState16Corr}
\epsilon = 1 - \frac{1-\cos 2 \pi 2^t \delta }{2^{2 t}} \sum_{\ell=-2^{p-1}+1 }^{2^{p-1}}  \frac{1}{1- \cos 2 \pi ( \delta - \ell/2^t ) } 
\ee
and if we define $ a = 2^t \delta $ and rearrange the cosine term in the summation we find
\be \label{eq:transformState16CorrCosec}
\epsilon = 1 - \frac{1-\cos 2 \pi a }{2^{2 t+1}} \sum_{\ell=-2^{p-1}+1 }^{2^{p-1}} \csc^2 \frac{\pi}{2^t} ( a - \ell )  .
\ee

Next, we demonstrate that the right hand side of Eq.~(\ref{eq:transformState16CorrCosec}) takes its maximum value at $ a = \frac{1}{2} $.  Since we know $ 0 \le a < 1 $, and since we expect the maximum value of $ \epsilon = \epsilon (a,t,p) $ to lie about midway between the two nearest states to generate the largest error, that is at $ a =1/2 $, we will substitute $ a = \frac{1}{2} + \Delta $, where $ \Delta \ll \frac{1}{2} $. To maximize $ \epsilon $ we need to minimize
\be \label{eq:transformState16CorrCosecDelta}
\cos 2 \pi \left ( \frac{1}{2}+\Delta \right )  \sum_{\ell=-2^{p-1}+1 }^{2^{p-1}} \csc^2 \frac{\pi}{2^t} \left ( \frac{1}{2} - \ell +\Delta \right )  ,
\ee
as a function of $ \Delta $. Expanding to quadratic order with a Taylor series, we seek to minimize
\be \label{eq:transformState16CorrCosecDeltaSecond}
\left ( 1-\pi^2 \Delta^2 + O(\Delta^4) \right ) \left ( c_0 + c_1 \Delta + c_2 \Delta^2 + c_3 \Delta^3 + O(\Delta^4) \right ) ,
\ee
where $ c_i $ are the coefficients of the Taylor expansion of cosecant$ ^2 $ in $ \Delta $.
We find by the odd symmetry of  the cotangent about $ \ell = \frac{1}{2} $ that
\be
c_1 =\frac{ 2 \pi}{2^t} \sum_{\ell=-2^{p-1}+1 }^{2^{p-1}} \cot \frac{ \pi}{2^t} ( \frac{1}{2} - \ell ) \csc^2 \frac{ \pi}{2^t} ( \frac{1}{2} - \ell ) = 0 ,
\ee
and so we just need to minimize 
\be
c_0 +(c_2 - c_0 \pi^2) \Delta^2  + O( \Delta^3) .
\ee
Differentiating, we see we have an extremum at $ \Delta = 0 $, and therefore $ \epsilon (a,t,p) $ has a maximum at $ a =1/2 $. 

Substituting $ a = \frac{1}{2} $ we obtain
\be \label{eq:transformState16CorrAFinal}
\epsilon \le 1 - \frac{2}{2^{2 t}} \sum_{\ell=-2^{p-1}+1 }^{2^{p-1}}  \frac{1}{1- \cos \frac{2 \pi}{2^t} ( \frac{1}{2} - \ell ) } .
\ee
We note that the summation is symmetrical about $ \ell = 1/2 $, and substituting $ t = p + s $, we obtain for our final result
\be \label{eq:transformState16CorrAFinal2}
\epsilon(s,p) = 1 - \frac{1}{2^{2(p+s)-2}} \sum_{\ell=1 }^{2^{p-1}}  \frac{1}{1- \cos \frac{\pi ( 2 \ell -1 )}{2^{(p+s)}}  }.
\ee
That is, given a desired accuracy of $ s $ bits, then if we add $ p $ more bits, we have a probability of success given by $ 1 - \epsilon $, of obtaining a measurement to at least $ s $ bits of accuracy.
Thus we have succeeded in deriving a best possible bound for the failure rate $ \epsilon = \epsilon(s,p)$.

\subsection*{Special Cases}

Numerical calculations show that $ \epsilon (t,p) $ quickly approaches its asymptotic value as $ t \rightarrow \infty $, and this limit gives a fairly accurate upper bound for $ \epsilon $, for $ t $ greater than about 10 qubits.
Using $ \cos x \ge 1 - \frac{x^2}{2} $ which is valid for all $ x $, and is accurate for $ x = O(1/2^t) $ as $ t \rightarrow \infty $,
\bea \label{eq:transformState16tlarge}
\epsilon & \le & 1 - \frac{4 }{2^{2 t}} \sum_{\ell=1 }^{2^{p-1}}  \frac{1}{1- (1 - \frac{1}{2} (\frac{\pi}{2^t} (  2\ell -1 ))^2) }  \nonumber \\
 & = & 1 - \frac{8}{\pi^2} \sum_{\ell=1 }^{2^{p-1}}  \frac{1}{ ( 2 \ell - 1 )^2  } .
\eea
An exact form for this can be found in terms of the trigamma function, being a special case of the polygamma function as shown in Abramowitz and Stegun \cite{Abramowitz:1964}, Eq.~6.4.5 :
\be \label{eq:transformState19a}
\epsilon \le  \frac{ 2}{\pi^2}  \psi'\left (\frac{1+2^p}{2}\right )
\ee
where $ \psi'(z) = \frac{d \psi}{dz} $ is the trigamma function, $ \psi(z) = \frac{\Gamma'(z)}{\Gamma(z)} $ is the digamma function, and $ \Gamma(z)= \int_0^{\infty} t^{z-1} \rme^{-t} dt $ is the standard gamma function.

\bigskip

Now considering the $ p \rightarrow \infty $ limit, which also includes the $ t \rightarrow \infty $ limit because $ t = p + s $, we can find an asymptotic form in the limit of large $ p $ also from \cite{Abramowitz:1964}, Eq.~6.4.12, namely
\be \label{eq:pinfinity}
\epsilon = \frac{4}{{\pi}^2 } 2^{-p} ,
\ee
which shows that the error rate drops off exponentially with $ p $ extra qubits.
The formula Eq.~(\ref{eq:pinfinity}) can be re-arranged to give 
\be \label{eq:pinfinityinv}
p_{\infty} = \left \lceil \log_2  \frac{2 \sqrt{2} }{{\pi}^2 \epsilon }  \right \rceil
\ee
which can be compared with the previous approximate formula shown in Eq.~(\ref{eq:transformState97}).

\bigskip

We have checked the new error formula through simulations, by running the phase estimation algorithm on a 2-dimensional rotation matrix, and undertaking a numerical search for the rotation angle that maximizes the error $ \epsilon $, which has confirmed Eq.~(\ref{eq:transformState16CorrAFinal2}) to six decimal places.

\section*{Discussion}

An exact formula is derived for the probability of error in the quantum phase estimation procedure, as shown in Eq.~(\ref{eq:transformState16CorrAFinal2}).  That is, to calculate ${ \phi }$ accurate to a required  ${ s }$ bits with
a given probability of success ${ 1 - \epsilon }$ we add $ p $ extra qubits, where $ p $ is given by Eq.~(\ref{eq:transformState16CorrAFinal2}).
If we have a large number of qubits then we can use Eq.~(\ref{eq:transformState19a}) valid at the $ t \rightarrow \infty $ limit. In the $ p \rightarrow \infty $ limit the asymptote is found as a simple exponential form Eq.~(\ref{eq:pinfinity}). 

The exact formula avoids overestimating the number of qubits actually required in order to achieve a given reliability for phase estimation and we have also found this formula to be useful in confirming the operation of classical simulators of the phase estimation procedure.

% You may title this section "Methods" or "Models". 
% "Models" is not a valid title for PLoS ONE authors. However, PLoS ONE
% authors may use "Analysis" 
%\section*{Materials and Methods}

% Do NOT remove this, even if you are not including acknowledgments
\section*{Acknowledgments}

Discussions with Anthony G. Williams and Sanjeev Naguleswaran
during the early stages of this work are gratefully acknowledged.

%\section*{References}
% The bibtex filename
\bibliography{precision}

%\section*{Figure Legends}

%\begin{figure}[!ht]
%\begin{center}
%%\includegraphics[width=4in]{figure_name.2.eps}
%\end{center}
%\caption{
%{\bf Bold the first sentence.}  Rest of figure 2  caption.  Caption 
%should be left justified, as specified by the options to the caption 
%package.
%}
%\label{Figure_label}
%\end{figure}

%\section*{Tables}
%\begin{table}[!ht]
%\caption{
%\bf{Table title}}
%\begin{tabular}{|c|c|c|}
%table information
%\end{tabular}
%\begin{flushleft}Table caption
%\end{flushleft}
%\label{tab:label}
% \end{table}

\end{document}